\documentclass[12pt]{article}
\usepackage{amsmath,amsthm}
\usepackage{graphicx,xcolor}
\begin{document}

	\author{Soumya Basu\thanks{%
			Department of Computer Science, Cornell University} \and David Easley\thanks{%
			Departments of Information Science and Economics, Cornell University} \and Maureen O'Hara\thanks{%
			College of Business, Cornell University}  \and Emin G{\"u}n Sirer\footnotemark[1]}
	\date{ December, 2018 }
	\title{Towards a Functional Fee Market for Cryptocurrencies\\
	}
	\maketitle
	
	\begin{abstract}
    Blockchain-based cryptocurrencies prioritize transactions based on their fees, creating a unique kind of fee market. Empirically, this market
    has failed to yield stable equilibria with predictable prices for desired levels of service. We argue that this is due to the absence
    of a dominant strategy equilibrium in the current fee mechanism. We propose an alternative fee setting mechanism that is inspired by generalized 
    second price auctions. The design of such a mechanism is challenging because
    miners can use any criteria for including transactions and can manipulate the results of the auction after seeing the proposed fees.
    Nonetheless, we show that our proposed protocol is free from manipulation as the number of users increases.
    We further show that, for a large number of users and miners, 
    the gain from manipulation is small for all parties. This results in users proposing fees that represent their true
    utility and lower variance of revenue for miners.
    Historical analysis shows that Bitcoin users could have saved \$272,528,000 USD in transaction fees while miners could have reduced 
    the variance of fee income by an average factor of $7.4$ times.
    \end{abstract}
	
	\pagebreak
	
\section{Introduction}

Almost all decentralized cryptocurrencies use the same basic mechanism to prioritize transactions. First, users attach a 
fee to each transaction. Then, miners choose the highest paying transactions to include in the blocks. Upon inclusion, 
the users pay the fee that was attached to their transaction. This mechanism plays a crucial role in these
cryptocurrencies as they subsidize miners to keep constructing the chain as well as ensuring that the capacity of the
network is being used efficiently. As the baseline subsidy to miners (the block reward) goes to zero and the
popularity of these cryptocurrencies increases, the income from transaction fees will take on an even 
more prominent role.

In practice, this mechanism has been very unstable. In Bitcoin, the average daily fee paid in June 2018 ranged from 
\$0.58 to \$6.85. In December 2017, during a period of heavy trading activity, the average daily fee ranged from
\$5.82 to \$61.44. For users, this volatility makes it very difficult to decide what fee to attach to a transaction and results in 
a poor user experience. Users who bid a low fee are dissatisfied by a long wait time, and the potential for their transaction to be dropped,
while users who bid too high may have overbid to get confirmed. Additionally, the unpredictability in fees drastically changes the mining rewards 
on a day-to-day basis. During June 2018, the total daily transaction fees earned by miners ranged from \$11,300 to 
\$1,348,000, while in December 2017, the total daily transaction fees ranged from \$2,039,984 to \$23,167,981. This makes it difficult 
for miners to make long term profitability plans.

While we use Bitcoin (BTC) as a case study for designing a functional fee market, our techniques can be fairly easily adapted to other decentralized
cryptocurrencies. Bitcoin is a decentralized system in which users can attach fees of any amount to a transaction and miners can include 
any such transactions in a block. Typically, miners put high fee transactions into the block and leave low ones out. So the fee a user offers can affect 
the waiting time that user experiences for his transaction to be recorded. However, as the rate at which blocks are
built and their size are both fixed by the protocol, fees do not result in more transactions being included and they 
affect average waiting times only if high fees discourage some users from participating in the system. As long as 
miner revenue induces enough miners to participate for security (total miner daily revenue ranged 
between \$10 million and \$17 million in June, 2018), these fees are a social waste.\footnote{See Easley, O'Hara and Basu [2017] and Huberman, Leshno and Moallemi [2017]
for analyses of transaction fees, the mining game and waiting times for users.} Indeed, most mining rewards in Bitcoin 
today are from the block reward and fees play only a small role in increasing miner revenue. As there is free 
entry into mining, expected miner profit is unchanged by fees. Thus, fees only change which transactions get processed
first.

Bitcoin's market mechanism to prioritize transactions is essentially a generalized first price auction
for space on the block. Here, the users act as bidders while miners are the auctioneers. First price auctions, 
or generalized first price auctions for multiple items, are problematic. In a first price auction for multiple,
identical items, the highest bidder pays his bid and gets the first item, the second highest bidder pays his bid and
gets the second item, and so on until either items or bidders are exhausted. These auctions do not have a dominant strategy 
equilibrium, and although Bayes Nash equilibria do exist and are efficient for generalized first price auctions for identical 
items with symmetric bidders, these equilibria are unlikely to occur in practice. In Bitcoin, this results in users not
revealing the full utility of their transaction in the fees, instead preferring to bid low at first and increase their
bid only if their transaction is taking too long to be confirmed. This strategic bidding leads to the observed fee instability in Bitcoin.

We propose a mechanism based on second price auctions, which
should perform better in practice. This mechanism has a simple goal: to enable more stable, predictable fees in
cryptocurrenties. This will disincentivize users from bidding strategically and prevent users from suffering from long wait times 
due to underbidding or overpaying for a transaction by overbidding. Miners will also gain a more predictable revenue
stream that will increase as the demand for block space increases. Additionally, since rational users will not underbid
strategically, miners can potentially earn more revenue from transaction fees as well. In total, our mechanism will
allow for a more stable, well behaved fee market.

While it may seem straightforward to apply the existing auction literature to cryptocurrencies, there are a few
characteristics of the domain that make it difficult to simply translate existing results. First, when a user sets a fee, 
they are allowed to view most of the previous bids made by other users. Also, a user is able to adjust the fee that they 
set, but they are only allowed to increase their bid in discrete amounts using a technique called child pays for parent.\footnote{Although users can observe the bids of other users and increase bids in response to their observations, we believe that analyzing the bidding game for a single block as a first-price, sealed-bid auction is appropriate for three reasons. First, users only have an approximate idea of what miners know so they don't get perfect information about the bids in the mempool. Second, users don't know when a block is going to be mined so they don't know when they need to be among the highest bidders. Third, the dispersion of bids we observe in Bitcoin is inconsistent with all users conditioning on others bids and only bidding enough to be among the successful users.} Further, miners,
acting as auctioneers, are allowed to place any transactions they want in a block including creating fee-paying transactions on the fly
to manipulate the fee mechanism. The protocol only controls what each user will pay for the transactions as well as the reward earned 
by the miner given the set of transactions chosen by the miner. This mechanism will be applied by all nodes and miners cannot deviate
from this mechanism in any way.
% OLD: {\color{red}{Miners and users are allowed to construct secret transactions that are not known to everyone, which allows miners to freely manipulate the auction mechanism that is used.}}
Miners and users are allowed to construct secret transactions that are not known to everyone, which allows miners to freely manipulate the auction mechanism that is used. There are two large classes of such transactions: fake bids that are submitted for the express purpose of manipulating the auction mechanism or transactions that have some intrinsic value to the miner. 
These limitations are inherent to most decentralized cryptocurrencies.
%does this mean more than just miners submitting fake transactions? We should expand on this or drop it. It sounds like we mean that they can interact outside of the protocol which we are ignoring/assuming away.
% Soumya: I think we handle the above assumptions, right?

Our mechanism is based on an adaptation of the generalized second price auction to cryptocurrencies. For our mechanism we show that as the number
of users increases, users' gain from bidding strategically converges to zero. This result shows that users have a nearly 
dominant strategy of bidding truthfully, especially as adoption increases. Additionally, we show that as adoption increases, the miner's
gain from manipulating the transactions they include in a block also converges to zero. We show this result both
empirically through simulations on real distributions and theoretically for certain distributions as well. Hence,
our mechanism is also resistant to a malicious miner manipulating the transactions they include in a block.

There are several recent papers that analyze the the current Bitcoin protocol, the games it induces and its efficiency or the lack thereof, and at least one paper that proposes an alternative protocol. Easley, O'Hara and Basu [2017] and Huberman, Leshno and Moallemi [2017] provide analyses of transaction fees, the mining game and waiting times for users in the current Bitcoin protocol. Huoy [2014] and Cong, He and Li [2018] provide analyses of the mining game. Boehm, Christin, and Moore [2015], Harvey [2016], Malinova and Park [2016], Raskin and Yermack [2016, 2017], and  Aune, Krellenstein, O'Hara, and Slama [2017] all analyze aspects of the Bitcoin environment. Rosenfeld [2011], Eyal and Sirer [2014], Gans and Halaburda [2015], and Gandel and Halaburda [2016] consider various design issues of the Bitcoin protocol mostly focusing on security rather than on the efficiency of the fee mechanism.

The most closely related work is the monopolistic miner protocol of Lavi, Sattath and Zohar [2017].\footnote{Yao [2018] provides proofs of conjectures from Lavi, Sattath and Zohar [2017] about the general incentive compatibility and profitability of their monopolistic miner protocol.} They propose a protocol in which the winning miner decides how many transactions to put into the block and charges all of them the lowest fee proposed by any transaction he placed in that block. There are fundamental differences between our approaches stemming from our goals and setup. Lavi, Sattah and Zohar [2017] assume a single monopolistic miner, and strive to maximize revenue from fees at a cost of lower social welfare. In contrast, our work explicitly targets maximizing social welfare, and operates under a model with many miners. In their system, the monopolistic miner is incentivized to leave transactions offering positive fees out of the block even if there is space in the block as including them reduces the uniform price he can charge. This, of course, maximizes miner revenue, but we believe that the first criterion for a viable protocol must be to use the blockchain efficiently, as otherwise users are discouraged from participation. Their non-manipulation result is stronger than the one we obtain from our mechanism since we only obtain declining gain from manipulation as the system grows, but it comes at a cost of lower social welfare and desirable metrics such as transaction throughput and latency. Finally, in both our protocol and the protocol proposed by Lavi, Sattah and Zohar [2017], users' incentive to behave strategically vanishes as the number of users grows.

In the rest of the paper, we first discuss the positive and negative aspects of using a multi-unit first price auction for slots on the 
Bitcoin blockchain. We then argue that auction theory and experience from the sponsored search search market suggests alternatives 
to the Bitcoin protocol. We adapt ideas from second price auctions and the VCG (Vickery-Clarke-Groves) procedure to the 
trustless, decentralized setting present in Bitcoin. Most importantly, in Bitcoin, prices charged to users cannot depend on bids from users who 
do not get on the blockchain and the miners cannot commit to not manipulate an auction. We show that a simple modification 
of standard results does apply to Bitcoin and we show how to create a superior protocol using those results.

\section{Auction Theory and Sponsored Search}

\begin{figure}
	\centering 
	\bf\large{Figure 1}\par\medskip
	\includegraphics[width=0.8\linewidth,keepaspectratio]{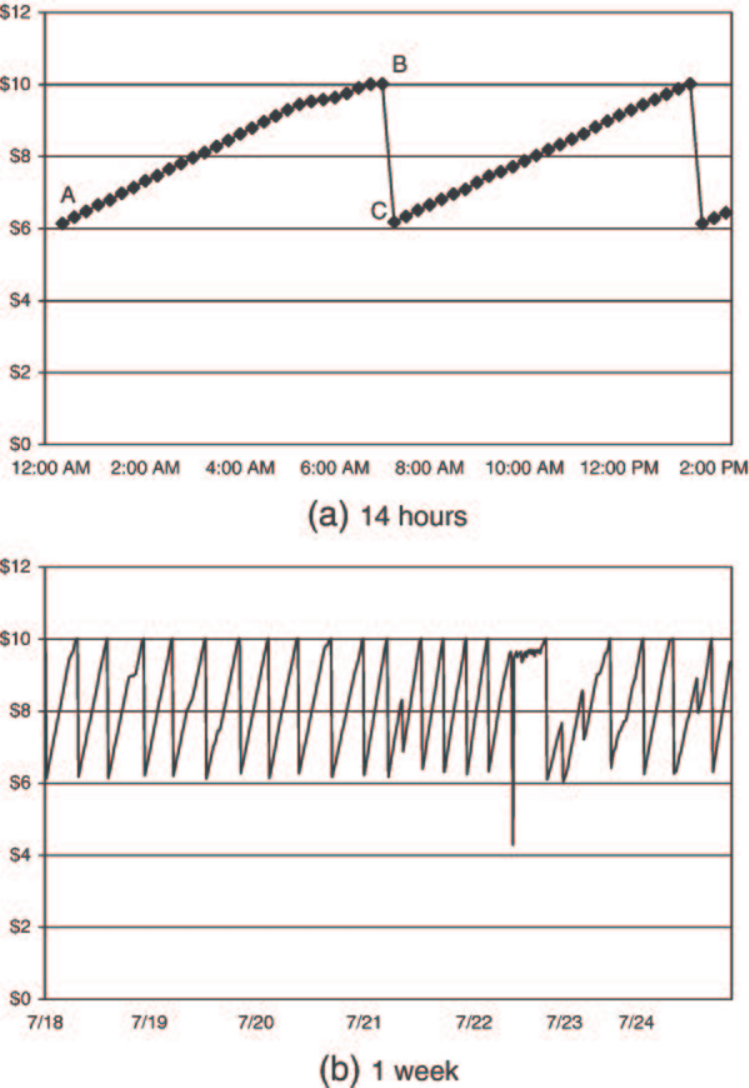}
	\caption{This figure illustrates the sawtooth behavior of bids on Overture when it was running a first price auction for ads to be displayed in response to searches using its search engine. The figure is reproduced from Edelman and Ostrovsky [2007].
	}
	\label{fig:overture}
\end{figure}

To understand the challenges and opportunities from auction design, it is useful to draw on the experiences of the sponsored search market.\footnote{See Easley and Kleinberg [2010], Chapter 15 for a discussion of the sponsored search market and references to the literature on sponsored search.} Overture, the first company to use keyword-based advertising, initially sold ads using a generalized first price auction. In the sponsored search market, advertising slots on the page that appears in response to a search term are sold to advertisers. Slots near the top of the page are preferred to ones down the page and all of these dominate those on the second page, and so on. Auctions are run frequently to determine whose ad appears where. Overture and its advertisers experienced  instability: bids in successive auctions would rise as those priced out in one auction tried to get into the next one; and then they would crash once bids reached levels that discouraged bidding at all. See Figure \ref{fig:overture}, reproduced from Edelman and Ostrovsky [2007], for an illustration of the saw-tooth behavior of bids in the Overture first price auction. 
Eventually, discouraged buyers quit and the auction was clearly producing less revenue than should be possible.

An important aspect of Google's subsequent success in the sponsored search market was based on its use of a superior auction form: GSP, Google's generalization of the single-unit second price auction to their multi-unit environment (multiple slots on the page). GSP does not have dominant strategies, but it is second-price-like, it is simple and it seems to work reasonably well. An alternative generalization of the single item second price auction to multiple items that does have dominant strategies is the Vickery-Clarke-Groves (VCG) procedure which forms the basis of the mechanism that Facebook uses to sell advertising.

Auction theory and the experience of the sponsored search market suggest that some generalization of the second price auction could be used to improve on the Bitcoin protocol. Before modifying a second price auction to fit the bitcoin environment it's useful to first set out our objectives in designing a protocol and then to describe how multi-unit second price auctions work. 

We have three objectives. First, the protocol should result in an efficient assignment of slots on each block to users of Bitcoin. So we want to assign slots to users with the highest values, leaving a user out of a block only if there is no user in the block who has a lower (true) value than the left-out user. An assignment with this property is called socially optimal. Second, we want the game induced by the protocol to incentivize non-strategic behavior. Ideally, we would like users' optimal bids (the fees they propose to pay) to be their true values for slots and we would like the miner building the block to have no profit motive for deviating from the ``rules of the auction.'' Third, we want optimal strategies to be simple and obvious. This last criterion is difficult to quantify, but a protocol that induces a game in which every participant (both users and miners) has (weakly) dominant truth-telling strategies surely satisfies it. 

In the standard auction environment, which does not fit the bitcoin environment perfectly, a generalized second price auction achieves these goals.\footnote{We call this a ``generalized'' second price auction as it is an auction for $K$ items at the $(K+1)$st highest bid rather than an auction for one item at the second highest bid. It is not the GSP procedure used by Google.} A generalized second price auction for $K$ identical items to be sold to bidders who each want at most one item works as follows. Bidders are asked to submit bids to the seller, or to the algorithm running the auction. The bidders who have submitted the $K$ highest bids each win an item and they all pay the $(K+1)$st highest bid. If the algorithm (auctioneer) can commit to this auction form, and if bidders private valuations for an item are iid draws from a fixed distribution, then it is a (weakly) dominant strategy for each bidder to bid truthfully---submit a bid equal to his value for an item. If such a dominant strategy did not exist, then although there may be a Bayes-Nash equilibrium, or multiple equilibria, they may be complex. Such complex equilibria make it unclear whether or not we should expect to see equilibrium bidding in practice. This auction form has another attractive feature---it guarantees that those who win are the ones who place the highest values on the items, so it results in an efficient assignment of items to bidders. It does not maximize the seller's expected payoff; doing that requires the seller to set a minimum bid at which an item will be sold and to reject any bids below that even if this results in unsold items.\footnote{This auction is equivalent to the VCG procedure. It is simpler to explain than VCG, but its optimality does depend on the items being identical. VCG does not require that restriction.} The following remark summarizes standard results about multi-unit auctions.
\bigskip

\noindent{\bf Remark:}
Suppose that each bidder wants at most one item and that bidders' values are drawn iid from a distribution on $[0,\bar{V}]$. Suppose also that the auctioneer has $K$ identical items for sale and can commit to an auction form. 
\begin{enumerate}
	\item If the auctioneer runs a generalized second price auction---the $K$ items are sold to the $K$ highest bidders at the $K+1$st highest bid---then it is a weakly dominant strategy for each bidder to bid his true value, and if each bidder follows this dominant strategy, the assignment induced by the auction is efficient
	\item If the number of bidders and the distribution of values is common knowledge, and the auctioneer runs a generalized first price auction---the $K$ items are sold to the $K$ highest bidders and each successful bidder pays his own bid---then there is a Bayes-Nash equilibrium of the game induced by the auction in which the assignment is efficient.
\end{enumerate}

\bigskip

For the environment described in the Remark, generalized first price auctions and generalized second price auctions both result in socially optimal assignments, so it's reasonable to ask why the second price-like auction does better than the first price-like auction in sponsored search. The difference is the complexity of the strategies, the knowledge required to find optimal strategies, and the notion of equilibrium. In the second price auction each bidder only needs to know his own value and the form of the auction. Bidding truthfully is optimal regardless of who the other bidders are or how they behave. This is not true in the first price auction. Here, the efficiency claim rests on the assumption that play can be described by a Bayes-Nash equilibrium in which each bidder is best responding to each other bidder. 

To illustrate the difference in these two auctions, it is useful to examine them in the simplest case in which there is a single item for sale to $N$ bidders with values, $V_i$, drawn iid from the uniform distribution on $[0,1]$. In a second price auction, it's weakly dominant for each bidder $i$ to simply bid his value $V_i$. In a first price auction there is an equilibrium in which the optimal strategy for a bidder with value $V_i$ is to bid $(\frac{N-1}{N})V_i$. This, of course, requires knowledge of the number of bidders; it depends on distribution of values being uniform; and, it is optimal only if all other bidders follow the same strategy. It does result in an efficient allocation because equilibrium bids are increasing in true values.\footnote{Both auctions also yield the same expected revenue for the seller (in equilibrium). For the example in the text a simple calculation shows this, but it's true much more generally according to Myerson's Revenue Equivalence Theorem, Myerson [1981].}

Improving on the current first price auction in the bitcoin environment is not as straightforward as simply adopting a generalized second price auction or the VCG procedure. Bitcoin is a trustless, decentralized system in which there is no central authority who can force all miners to commit to acting as if they are the auctioneer in a generalized second price auction. The fee setting game is not in fact an auction and there is no trusted auctioneer. So any protocol has to take into account the incentives of the miners to follow the ``rules'' of the auction rather than to manipulate it.
Second, only the miner knows the transactions and their attached fees in his mempool. Once he writes transactions to the blockchain the details of those transactions are known, but details of the transactions he leaves out are not known---and the protocol cannot credibly call for payments that depend on those left-out transactions. Even worse, there is a lag between when the block is constructed and when the block is mined where new bids can appear. This makes it impossible for anyone but the miner to know the bids that are involved when constructing the block. So the generalized second price auction, selling $K$ items at the $K+1$st highest bid, is not feasible as the uniform price is not publicly observable. The incentives the miner has to manipulate the auction are even more problematic and we discuss them next.

The winning miner has no mechanism allowing him to commit to an auction form.\footnote{Akbarpour and Li [2018] provide an analysis of mechanisms in which the seller can deviate from the rules of the auction. In this case, the mechanism has to be incentive compatible for the seller. They show that a first price auction is the only credible static auction. Essentially, an auctioneer could announce a different auction, such as a second price auction, but then once bids are received, he can submit a false second highest bid just below the actual highest bid---turning the auction into a first price auction. Credibility also matters for our analysis as our miner can submit own bids; but our environment differs as there are multiple items for sale, the mechanism can impose some constraints on the miners, and, most importantly, miners revenue can depend on the fees generated by a sequence of blocks determined by the protocol.}  Most importantly, the miner can also act as a user and include his own transactions into the block he is mining, moving money from one of his wallets to another with whatever fee he chooses after observing the fees offered by users. All identities on the blockchain (miners, users, etc) are uniquely identified by a cryptographic key. Thus, it is cheap to create a new identity, but hard to assume the identity of another person. This makes it difficult to enforce roles for each participant since it is possible for a miner to also impersonate other, arbitrarily many, identities which all act as ``users''.

This ability to also act as a ``user'' or many ``users'' allows a miner who earns the revenue generated by the block to introduce first-price-like features into a supposedly second price auction at zero cost to himself as he pays the fee on his fictitious transaction to himself. To see this in the simplest case, suppose that there is only one transaction in the block. The miner can include a fictitious transaction paying a fee equal to the highest offered real fee, ensuring that the user with the highest offered fee wins and pays that fee. This makes the single item, ``second price'' auction with a strategic auctioneer effectively a first price auction. So bidders should place first price bids and in equilibrium, we should see a first price outcome. Nonetheless, we show that with multiple bidders and multiple slots on the block, a generalization of the second price auction can be useful. 

The explanation above applies to a static (one-shot) auction. Neither the sponsored search market nor the blockchain game are static. In sponsored search, auctions are run frequently and an advertiser who does not win now can change his bid and perhaps win in a subsequent auction. In the blockchain game, a user who does not get onto the current block can revise his fee offer and, after waiting longer, perhaps get onto the next block. Most of the sponsored search literature ignores this dynamic feature although there is recent work on dynamic sponsored search auctions.\footnote{See, for example, Edelman and Schwarz [2010].} We will take waiting time into account indirectly, through users differing values for a spot on the block and we will make use of a sequence of blocks, but we will analyze the users' blockchain game only one block at a time. 

\section{Model}

We consider a generic model applicable to a broad range of cryptocurrencies. Cryptocurrencies construct blocks with a fixed number of slots, $K$, that can be filled with users' transactions selected by the miner who is building this block.\footnote{For ease of exposition, our model uses the terminology used in popular proof of work cryptocurrencies, such as Bitcoin or Ethereum. Most cryptocurrencies operate on batches of transactions analogous to blocks, and transaction priority is decided by user's bids.} We assume that there are $N>K$ users. Some of these users have transactions waiting in the mempool at the time the current block is being constructed and others are absent. We assume that the probability that a user has a transaction in the mempool is $\delta$ where $1>\delta>0$ and users are drawn iid. The users waiting in the mempool have heterogeneous values for having their transaction recorded to the blockchain. These values are denoted $V_i$ and they are drawn iid according to a continuous density $g$ on $[0,\bar{V}]$ with $g(V)\geq \epsilon >0$ for all $V\in[0,\bar{V}]$. A user who is not included in the current block receives no reward from the current block.\footnote{For simplicity of exposition, we treat all transactions as taking the same amount of space on the blockchain. In practice, the fees we discuss are normalized in some way. For example, in Bitcoin, this would be the fee per byte. Dependent transactions, such as child pays for parent, can be handled by charging the average fee for both transactions.}

All users have attached transaction fees (bids) to their transactions denoted by $f_i\in[0,\bar{V}]$. We assume that distributions of users and values are common knowledge. We model users as selecting bids after knowing their own value and knowing how many users are active, but without knowing the values or bids of other users. Miners select which transactions to put into the block after seeing the bids attached to those transactions. We consider only blocks for which the number of active users is greater than the number of slots ($K$).

Users can attach any fee they like, or no fee, to their transaction. If the miner of the current block places a user's transaction in the block, then the miner keeps the fee.  A profit maximizing miner clearly selects the $K$ highest bids, or all bids if there are less than $K$ bids, places those transactions on the block, and earns those bids. The miner has no incentive to manipulate by entering fictitious transactions as by doing so, he simply removes real transactions and the fees they generate without changing the fees paid by other transactions. So our model of the current protocol is in fact equivalent to a generalized first price auction run by an auctioneer who can commit to running this auction type.
\bigskip

\noindent{\bf Remark:} Our model of bidding for slots in current cryptocurrency protocols induces a generalized first price auction for the $K$ slots on the current block. The users with the $K$ highest bids (proposed fees) win slots on the block and each winning user pays his bid (fee).

\bigskip

Thus, from the users' point of view the fee setting game is a generalized first price auction for $K$ identical items. This auction has a symmetric Bayes-Nash equilibrium in which equilibrium bids are increasing in values and so, in this equilibrium, the highest value transactions are placed on the block.\footnote{Symmetry of users matters for this claim. With asymmetric distributions, equilibria are not symmetric, and efficiency need not occur. It also depends on our assumption that each user has only one transaction in the mempool. If users are interested in multiple slots on the block, then efficiency can also be lost.} In this equilibrium, realized social surplus, defined as the sum of the values of users whose transactions are placed in the block, is maximized. Miner revenue and the total payment by users net out and so do not affect social surplus. Of course, this is an interim notion of social surplus as it takes the winning miner as fixed and does not consider the cost of the mining industry. We address these concerns later. Proofs of Results are included in the Appendix.
\bigskip

\noindent{\bf Result 1:} Our model of current cryptocurrency protocols applied to a single block of size $K$ induces a game which has a symmetric Bayes-Nash equilibrium which is efficient.

\bigskip

If miners could commit to an auction form and the protocol could use all bids to determine the assignment and payments, then it could be modified to induce a generalized second price auction for the $K$ slots. In this auction, the $K$ highest bidders would have their transactions placed on the block and they would pay a uniform price equal to the $K+1$st highest bid. In this auction it is a (weakly) dominant strategy for each bidder to bid his true value. To see this, note that each bidders' bid only affects whether or not his transaction is placed on the block; it does not affect the price he will pay if he gets on the block, as that price is the bid of a bidder who is not successful. A bidder wants to be on the block if the price is no more than his value and he does not want to be on the block otherwise. A bid equal to his true value ensures this. Most importantly, note that this reasoning does not depend on how many other bidders are present or on what they do; so bidding truthfully is a dominant strategy. Bidders using their dominant strategies result in maximum realized social surplus. 
\bigskip

\noindent{\bf Result 2:} If miners could commit to use a generalized second price auction and the protocol could use all bids to determine the outcome, then the protocol could be modified to induce a generalized second price auction. In this auction, truth-telling would be a (weakly) dominant strategy for users and the assignment would be efficient.

\bigskip

Because miners cannot commit to an auction form and the $K+1$st bid is not observable, this generalized second price auction is not feasible. Most importantly, the miner's ability to submit a fake transaction after observing the fees in the mempool destroys the ``truth-telling is a dominant strategy'' result.\footnote{If the miner could only submit fictitious transactions before observing the fees in the mempool, then truth-telling would remain a dominant strategy for users. From the point of view of users the miner is just acting as another user or users and this has no effect on any user's incentive to bid truthfully.} For a block with $K$ slots the optimal fictitious transaction(s) by the miner are not as simple as just matching the $K$th highest bid; the equivalent of matching the highest bid in one-unit ``second price auction'' thereby turning it into a first price auction. For example, suppose that $K=2$ and there are three transactions in the mempool with attached fees of $f_1>f_2>f_3$. Then there are two possible manipulations by the miner: (1) Insert a transaction with fee equal to $f_2$ and earn total fee of $2f_2$, or (2) Insert a transaction with fee equal to $f_1$ and earn total fee of $f_1$. Manipulation (1) is better than the generalized second price auction (for the miner) and (2) is better than (1) if $f_1/2>f_2 $. Thus, the miner always has an incentive to manipulate unless there are $K+1$ bids of equal highest value. 

\section{Proposed Mechanism} 

To solve both the inability to use the $K+1$st highest bid and the miner's incentive to manipulate, we propose the following protocol. In any cryptocurrency, generating identities is very cheap so miners are able to create blocks under new identities that they control. This protocol will produce the same results for a miner independent of which identities they mine under, removing this concern from our analysis.
%%MO can we expand on what the sentence above means?

\bigskip
\noindent{\bf Protocol} 
\begin{enumerate}
	\item Label blocks by $b=1, 2,\dots$.
	\item Users who want a transaction recorded in block $b$ can attach fees to their transactions. Denote the fee attached by user $i$ by $f_i$.
	\item Users whose transactions are included in block $b$ each pay the minimum fee proposed by any user whose transaction is included in block $b$. The total paid by the these users in block $b$ is the revenue generated by block $b$.
	\item The miner who builds block $b$ is paid the average revenue generated by the $B$ most recently mined blocks, including block $b$, if and only if the miner fills block $b$. Otherwise, the block is not included in the blockchain.
	\begin{itemize}
		\item  A block is defined to be filled if it contains $K$ transactions or if the miner pays a \emph{fill penalty}. The necessary fill level, $K$, is a parameter which can be chosen to be some fraction, say $80\%$, of the capacity of a block. The fill penalty is defined to be the difference between $K$ and the number of transactions in the block times the fee paid by each transaction. A miner can also avoid paying the fill penalty by declaring that there were not enough transactions in the mempool to fill the block, in which case each user is charged the minimum allowable fee for a transaction to be included in the mempool.
		\item A minimum fee required for a transaction to be considered can be included by declaring that transactions are not in the mempool if the proposed fee is below that minimum level.\footnote{Including a minimum bid is standard in sponsored search auctions. For example the minimum bid that Google currently uses is one-cent per click.}
		\item The miner has the option to fill the block to capacity with transactions, but only $K$ of them are priced using this auction mechanism. The other transactions are charged no fees for being included in this block.
	\end{itemize}
\end{enumerate}

The reward for a miner's payment from transaction fees must incentivize the miner to not manipulate the contents of the block and to place the highest paying transactions from the mempool into the block. To deal with both concerns, we set the miner's reward to be the average revenue generated over the last $B$ blocks, including the block that the miner has just mined.\footnote{The fill penalty is considered part of the reward for a block.}

The miner is incentivized to place the $K$ highest value transactions into the block for two key reasons. First, the miner is receiving a fraction of the rewards in the block that they recently mined. Thus, they are directly incentivized to maximize the rewards in this block. Additionally, miners are paid in the underlying cryptocurrency and they typically maintain a position in the cryptocurrency. So, they have an incentive to make the cryptocurrency successful, which incentivizes them to put high value transactions into the block; that is, to take the $K$ highest fee transactions.\footnote{Note that which transactions (with fees above the minimum fee over all transactions in the block) the miner places in the block (provided that he is not a party to those transactions) has no effect on his revenue. So we are assuming that if he is otherwise indifferent, concerns about the success of the cryptocurrency leads him to place the highest value transactions on the block.}

The protocol incentivizes the miner to not manipulate the transactions in a block due to the fill penalty. The optimal manipulation for a miner is to insert fake bids that are equal to the minimum bid from some user that the miner wants to include in the block. However, the fill penalty is equal to the fees that the miner would pay to perform this manipulation. Since larger blocks are more likely to get forked and excluded from the blockchain due to random chance,\footnote{Forking occurs when two blocks are mined that cannot both be included in the blockchain (e.g. when they have the same height). Any block that is mined has a chance of being forked, but larger blocks have a higher chance of being forked. The probability of a fork occurring depends on the particular chain's properties.}
%%MO we should at least include a footnote to explain comment about forking
a miner is incentivized to prefer smaller blocks, all else being equal. Additionally, each fake transaction from the miner will require the miner to pay the associated fee. Since each block's reward is computed over the past $B$ blocks, the fees collected from a particular block are actually spread over the next $B$ blocks (including itself). Thus, the miner can only get a fraction of her reward back on expectation.\footnote{This fraction is precisely equal to the fraction of mining power that this miner has.} Exactly how many blocks $B$ to average over is an empirical question that will vary on a per-blockchain basis.

Similarly, the protocol also incentivizes miners to place transactions that have value to the miner using the space in the block not under the auction mechanism. This enables the miner to capture the full value of the transaction and include it in the blockchain without paying other miners a fee. This includes transactions that have business value (e.g. payments to members in their mining pool) or side payments from users to include their transaction in the block.

Assuming that miners place the $K$ highest fee transactions on the block, and that users whose transactions are placed on the block all pay the $K$th highest fee, it is no longer a dominant strategy for users to bid truthfully as the $K$th highest bidder's (who could ex ante be anyone) bid affects the price he pays. So users have an incentive to manipulate, however, that incentive is small if there is a large number of users. To see this, suppose that all other users bid truthfully. Let $V_{K-1}$ and $V_K$ be the $K-1$st highest bid of others and the $K$th highest bid of others. If a user's value is below $V_K$ there is no possible gain from bidding strategically as the price will be greater than the user's value for any bid. There is a potential profit from strategic bidding only if the user's value is greater than $V_K$ and this gain is bounded by $V_{K-1} - V_K$. So given a user $i$ with value $V_i$ the gain to strategic bidding is bounded by $E[(V_{K-1}-V_K)]$ which converges to $0$  in the number of users. For example, with draws of user values according to the uniform distribution on $[0,1]$ and $A$ active users, the upper bound on gain is $1/A$ and with the exponential with parameter $\lambda$, it is $1/\lambda (A-K+1)$. That is, with a large number of users, the potential gain to strategic user behavior is small and it seems likely that users will instead follow the simpler, nearly optimal, strategy of truthful bidding.

\bigskip

\noindent{\bf Result 3:} For the protocol and model above:
\begin{itemize}
	\item Truthful bidding is not a dominant strategy for users.
	\item If all other users bid truthfully the expected gain to any bidder from strategic bidding converges to $0$ as the number of users diverges.
\end{itemize}

\bigskip

\section{Large Number of Users}

If the number of users is large, the protocol can be made simpler without much chance of harm. Suppose that the protocol is modified to pay the miner of block $b$ the revenue generated by block $b$. A miner could now manipulate by inserting a fictitious transaction with a fee equal to any of the $K$ highest fees offered.\footnote{Inserting a fee between two existing fees is clearly dominated by making the fictitious fee equal to the higher of two nearby fees. Inserting multiple fictitious fees above the $K$th highest fee knocks some number of transactions out of the block and sets the price at the lowest fee remaining in block and so is equivalent to a single fictitious fee strategy.} Relabeling the $K$ highest fees from highest to lowest, they are $f_1, f_2, \dots, f_K$. For a miner to not manipulate, we need the revenue generated from $K$ transactions at the $K$th highest bid to be greater than the revenue generated from any smaller number of transactions $n$ at the $n$th highest bid, i.e. $Kf_K>(K-1)f_{K-1}$, $Kf_K>(K-2)f_{K-2}$, and so on. This clearly holds if it holds sequentially, i.e. $Kf_K>(K-1)f_{K-1}$, $(K-1)f_{K-1}>(K-2)f_{K-2}$, and so on. This second collection of inequalities can be written as $nf_n>(n-1)f_{n-1}$ for each $n=2, \dots, K$. Or $f_n>(n-1)(f_{n-1}-f_n)$ for each $n=2, \dots, K$. If users bid truthfully, then as the number of users diverges, the left-hand side of this inequality converges to $\bar{V}$ for any fixed $n$ and the right hand side converges to $0$. So the miners' gain from manipulation vanishes as the number of active users grows.\footnote{Note that inserting a fictitious transaction with fee greater than the $K$th highest real fee is equivalent to choosing to not fill the $K$ slots on the block. So this argument also covers the potential manipulation of restricting the supply of slots.}

\bigskip

\noindent{\bf Result 4:} Suppose there are $K$ slots in the block. For both the generalized first price auction and the generalized second price auction with uniform price equal to the $K$th highest bid and for any Bayes-Nash equilibria of the these auctions, the ratio of realized social surplus to maximal social surplus and the ratio of realized miner revenue to maximum miner revenue converges to one with probability one.

\bigskip

So with a sufficiently large number of users, both miners and users have little incentive to manipulate the generalized second price auction (with price equal to the $K$th highest fee). 

\section {Analysis}

Our analysis aims to answer three key questions.
First, we use simulation in order to understand the miner's incentive to manipulate the auction and deviate from our proposed protocol.
Second, we analyze bids during a period where block space was scarce in Bitcoin and Ethereum to estimate how much users are overpaying using a first price mechanism.
Finally, we use the same bids to estimate the reduction in variance from fee revenue for miners.

\subsection{Miner Manipulation}

To help understand a miner's incentive to manipulate our proposed protocol, we ran a series of simulations. Our simulations show that the gain to miners from manipulation decreases as the number of miners increases or as the number of blocks we average over in determining miner fees increases. Overall, our results support our claim that, for reasonable parameters, even optimal manipulation by miners does not pay.

Our simulations take as parameters the number of transactions (N) that are in the mempool, the number of transactions (K) that can be included in a block, the number of miners (M) and the number of blocks that the transaction fees are averaged over (B).

To run the simulations, we first draw user bids from a power law distribution with a median of 2 cents and a mean of 10 cents, which is similar the actual transaction fee distribution that appeared on the blockchain in July 2018. The miner then chooses $j$ real transactions to include in a block (where $j \leq K$) and then fills the rest of the block with fake transactions. The total fee generated by this block is $j$ times the $j$th highest bid from the user distribution, which we denote as $b_j$. Since the manipulating miner solved this block, he receives $\frac{b_j}{B}$ as a reward from fees on this block. 

A miner with $\frac{1}{M}$ of the hashpower on expectation will receive
$\frac{b_j (B-1)}{MB}$ in fees because this miner is expected to mine $\frac{1}{M}$ of the  other $B-1$ blocks that the transaction fees are averaged over. Finally, we subtract the amount that the miner paid to insert fake transactions and manipulate the fees, which is $(K-j)b_j$. We then take the maximum over all possible values of $j$ and define the gain as the maximum value
minus the miner revenue when $j = K$ and the miner has been honest---not inserted any fake transactions into the block. We average this over 1000 trials to compute the total gain. In all of our results, we keep the number of transactions in a block, $K$, fixed at 2000.

\begin{figure}
	\centering 
	\bf\large{Figure 2}
	\includegraphics[width=1.1\linewidth]{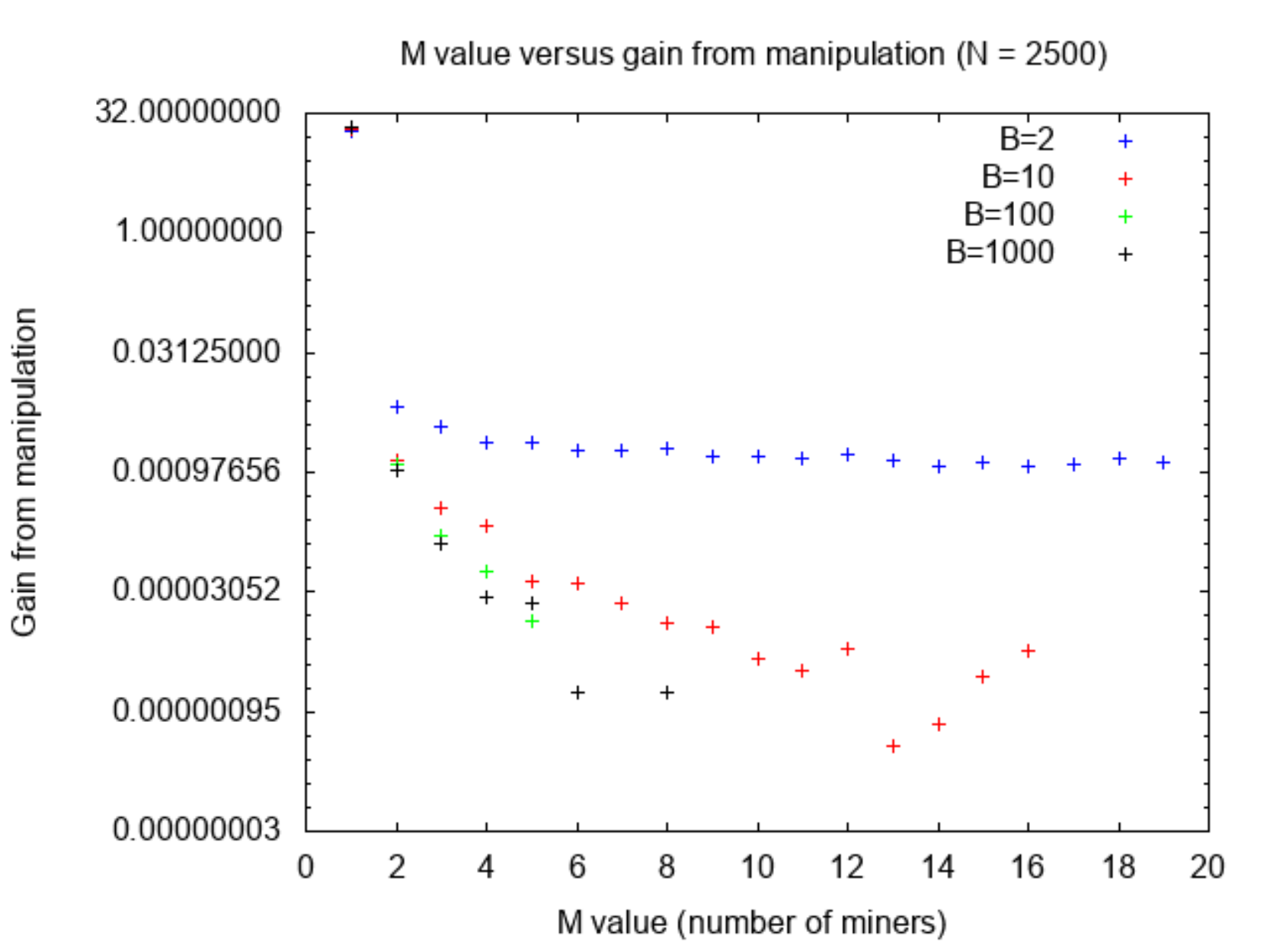}
	\caption{This figure illustrates the change in a miner's gain from manipulation as the number of miners increases. We keep the mempool size fixed at 2000 and examine this gain as we vary the number of blocks we spread the reward over (B). 
	The simulation shows that increasing the number of miners decreases the gain from manipulation as long as we average over enough blocks.	
	}
	\label{fig:miners}
\end{figure}

Figure \ref{fig:miners} shows miner's gain from optimal manipulation as the number of miners increases for various levels of the number of blocks we average over. The gain from manipulation declines as the number of miners increases and this decline is most pronounced if the number of blocks averaged over is large. This occurs because as the number of miners increases the probability of any individual miner winning a block declines and payment is averaged over (B) blocks while the cost of manipulation in a block does not depend on the number of miners or the number of blocks averaged over.

\begin{figure}
	\centering
	\bf\large{Figure 3}
	\includegraphics[width=1.1\linewidth]{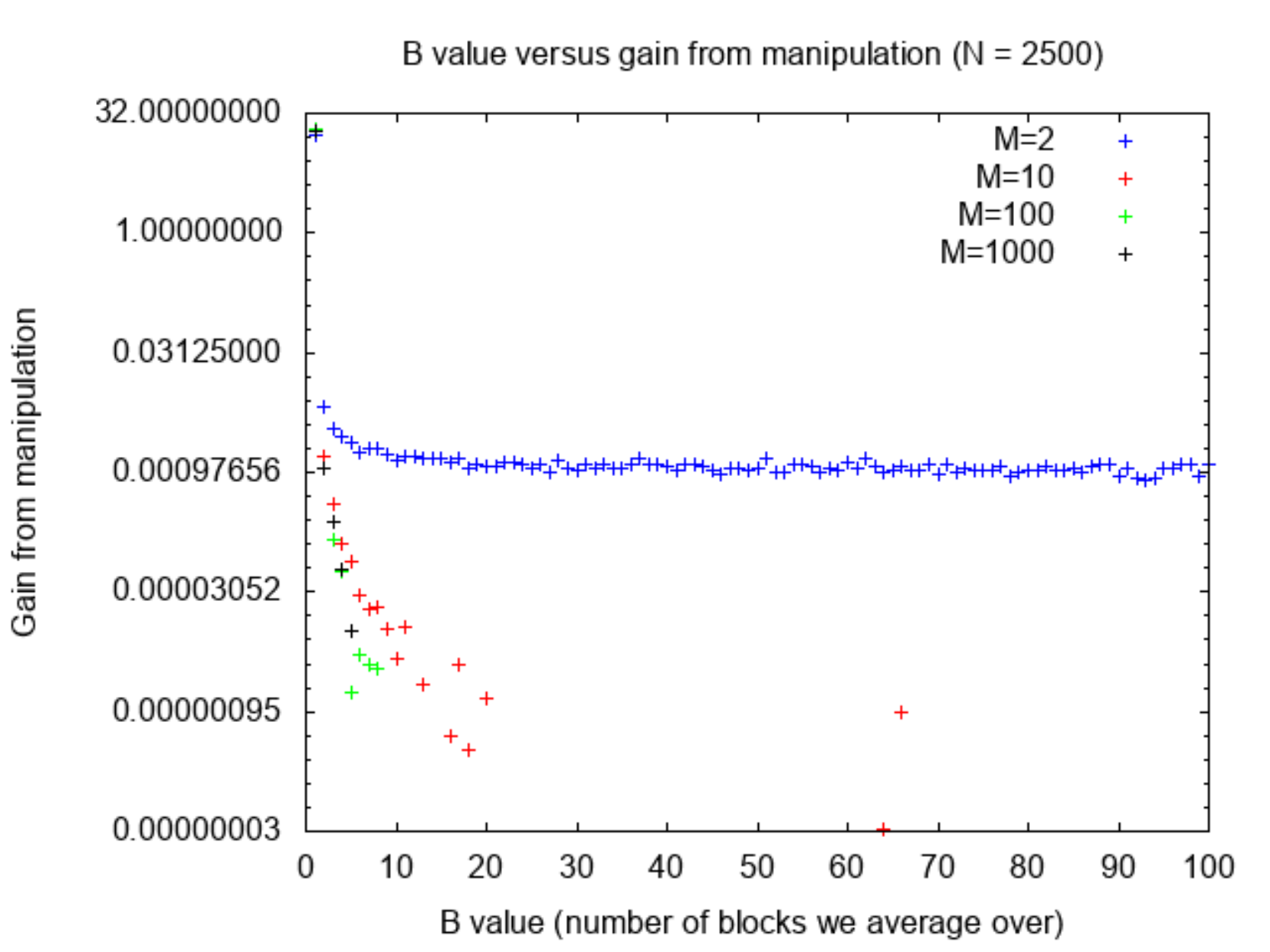}
	\caption{This figure illustrates the change in a miner's gain from manipulation as the number of blocks we average over increases. The number of transactions in the mempool is kept fixed at 2000. The simulation shows that increasing the number of blocks we average over decreases the gain from manipulation as long as there are enough miners.	}
	\label{fig:blocks}
\end{figure}

Figure \ref{fig:blocks} shows that the gain from optimal manipulation declines as the number of blocks averaged over increases and that it is uniformly lower for high numbers of miners. This occurs because the cost of manipulation is fixed and the expected gain declines as more blocks are averaged over.

Our simulation results show that miners do not gain much from manipulation, even if they do it optimally, for reasonable numbers of miners, users and blocks averaged over. Miner revenue ranges from \$15 to \$40 over these simulations, so the gain from manipulation relative to total revenue is very small for reasonable numbers of miners and blocks averaged over. Of course, this does not take into account the miners incentive to make Bitcoin succeed so as to maintain the value of their Bitcoin holdings and the ongoing value of the mining operation. We believe that taken together these results argue that miners are not likely to manipulate our proposed protocol.

\subsection{Blockchain Bid Analysis}

To help understand the concrete benefits of our proposal, we analyze the actual bids that appeared on the blockchain during a period of high demand in Bitcoin (December 2017).
Note that Bitcoin, during this time interval, is still operating with a protocol equivalent to a generalized first price auction.
Thus, the bids that we analyze are likely lower than the utility of the users.

\begin{figure}
	\centering
	\bf\large{Figure 4}
	\includegraphics[width=1.1\linewidth]{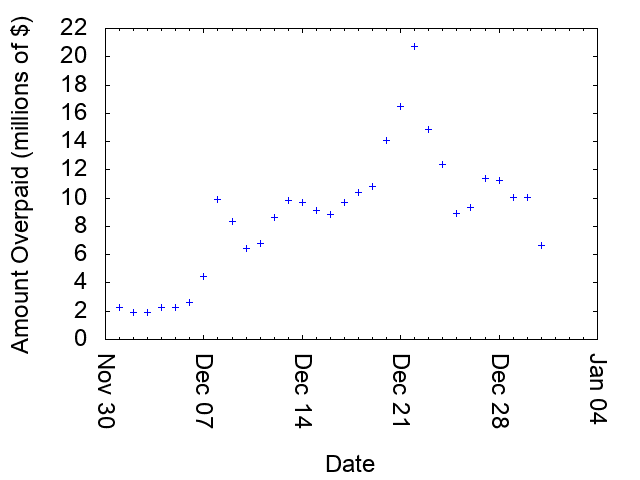}
	\caption{This figure illustrates how much (in millions of USD) that users could have saved if the transaction fees were using our proposed scheme rather than the generalized first price auction in Bitcoin in December 2017. We see that the daily savings is quite large and this is roughly equivalent to the social waste of the fee bidding scheme.}
	\label{fig:overpayment}
\end{figure}

\begin{figure}
	\centering
	\bf\large{Figure 5}
	\includegraphics[width=1.1\linewidth]{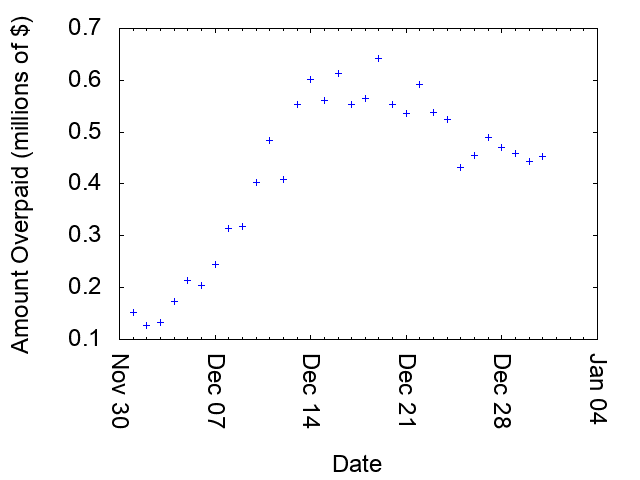}
	\caption{This figure illustrates how much (in millions of USD) that users could have saved if the transactions were using our proposed scheme rather than Ethereum's current scheme in December 2017. As Ethereum has higher throughput than Bitcoin, the savings are not as high.}
	\label{fig:ethoverpayment}
\end{figure}

We first examine how much clients could save using the auction mechanism that we propose in this paper.
To apply our mechanism to each day, we take all of the blocks that were mined that day and look at the transactions in each block.
We then calculate the fee per byte for each transaction to normalize the fees paid, and then have every transaction pay the smallest fee per byte that appears on the block per our mechanism.
Then, we plot the difference between the actual fees paid and the fees that users would pay under our mechanism.
Figure~\ref{fig:overpayment} shows that the difference in Bitcoin is quite substantial and that under times of high demand, there is a significant amount of social waste generated by the generalized first price auction.
Figure~\ref{fig:ethoverpayment} shows a similar trend in Ethereum, but the dollar amounts are significantly smaller, which is to be expected as Ethereum has a higher processing capacity than Bitcoin.

\begin{figure}
	\centering
	\bf\large{Figure 6}
	\includegraphics[width=1.1\linewidth]{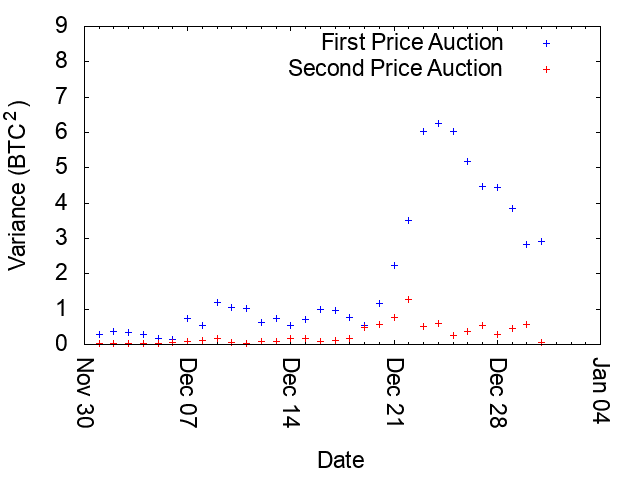}
	\caption{This figure illustrates the variance (in $BTC^2$) of the payouts from transaction fees to each miner for each day in December 2017. We see that the first price auction has a significantly higher variance
	than the second price auction mechanism, resulting in less stable payouts for miners.}
	\label{fig:variances}
\end{figure}

\begin{figure}
	\centering
	\bf\large{Figure 7}
	\includegraphics[width=1.1\linewidth]{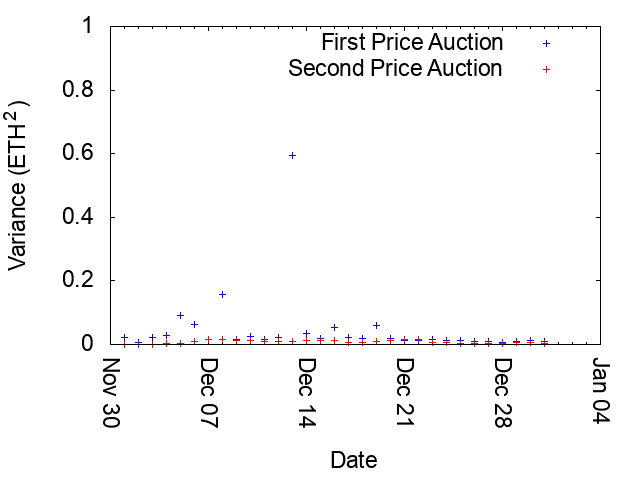}
	\caption{This figure illustrates the variance (in $ETH^2$) of the payouts from transactions fees to each miner for each day in December 2017. We see that the first price auction mechanism has a significantly higher variance than the second price mechanism.}
	\label{fig:ethvariances}
\end{figure}

Our mechanism has the potential to improve predictability by reducing variance.
To quantify this, we use the same calculation as before to obtain the fees that miners would receive under our mechanism.
This time, we take the variance of the transaction fees on each block using the second price mechanism and the current first price mechanism.
Figure~\ref{fig:variances} shows that the variance is lower in Bitcoin when using the second price mechanism, by up to a factor of $20$ on some days.
Figure~\ref{fig:ethvariances} shows that the same trend holds in Ethereum as well.
Additionally, the results in Figure~\ref{fig:variances} and Figure~\ref{fig:ethvariances} both assume that each block's payment only goes to one miner.
In our mechanism, payouts are averaged over $B$ blocks, which would further decrease the variance by an additional factor of $B^2$.

Our bid analysis shows that, using our proposed protocol, we are able to avoid social waste for users and decrease miner payout variability.
So, our mechanism is likely to perform well in practice, especially during periods of high demand.

\section{Conclusion}

Cryptocurrencies cannot go mainstream if constructing a transaction imposes a cognitive load or requires
complex strategic bidding behavior. We show that the fee mechanism currently used in a variety of coins encourages users to employ complex bidding strategies. We then present an alternative that obviates this need and offers a more stable, predictable fee market.

Both the generalized first price and generalized second price auctions work well in equilibrium if there are a large number of users relative to the capacity of blocks. So why is the generalized second price auction preferable? In the generalized second price auction the transaction fee offered by a user only affects what a successful user pays if the user has the (unique) $K$th highest bid. Otherwise the fee only affects whether the user is in the block or not in it. So the gain to strategic bidding is small if there are many users. But with a first price auction every user pays the fee that he offers if his transaction is in the block. Here, strategic bidding is inescapable although its gain does converge to zero as the number of users grows. This robustness of the second price procedure makes it more desirable than the first price procedure even if the two would have similar surpluses in equilibrium. 

How the choice of procedure (first or second price) affects miner revenue from fees is unclear. In equilibrium, there is revenue equivalence; the two procedures should produce approximately the same expected revenue. Empirically, we see that the miner revenue will have a lower variance payout under the second price procedure. But what happens to the actual payout when the game is played by real users and miners is unclear, at least in part because of the non-robustness of the first price procedure. 

Finally, we note that our analysis applies to proof-of-work protocols such as those used in Bitcoin, Ethereum and many others. Alternative protocols are being considered and used in a variety of different digital currencies. Most notably, Ethereum is considering a switch to proof of stake. Regardless of the protocol, cryptocurrencies will need to prioritize transactions somehow. Most cryptocurrencies charge fees to use the network and induce a first price auction. Thus, they face the same problems described above.

\section{Appendix}

\noindent{\bf Proof of Result 1:} By the Remark in the text this result is equivalent to showing that a discriminatory, multi-unit, first price auction has a symmetric Bayes-Nash equilibrium in which bids are increasing in values. This is a standard result, see Weber [1983] and Milgrom [1985].

\bigskip

\noindent{\bf Proof of Result 2:}  The modification yields a multi-unit, second price auction. The result that this auction has weakly dominant strategies and an efficient equilibrium is a standard result, see Weber [1983] and Milgrom [1985]. 

We provide a simple, direct proof of this claim as we use the logic elsewhere in the paper. Consider bidder $i$ with value $v_i$. We need to show that bidding more than $v_i$ or less than $v_i$ cannot increase the profit of bidder $i$. 

Consider a bid $b_i>v_i$. Bidder $i$'s bid only affects whether he wins or loses the auction; it does not affect the price he pays conditional on winning. So this high bid only changes the payoff to $i$ if bidder $i$ would not win with a bid of $v_i$ and would win with a bid of $b_i$. That is, only if $b_i>v^K>v_i$, where $v^K$ is the Kth lowest bid of the other bidders. In this case $i$ wins with a bid of $b_i$, but pays $v^K>v_i$ as $v^K$ is now the K+1 st highest bid. So high bidding reduces $i$'s payoff.

Alternatively suppose that $i$ bids $b_i<v_i$. This only affects $i$'s payoff if $i$ would have won with a bid of $v_i$ and does not win with a bid of $b_i$. That is, only if $v_i>v^K>b_i$. In this case $i$ would have won with a bid of $v_i$ and paid $v^K<v_i$ and does not win with a bid of $b_i$. So a low bid also reduces $i$'s payoff. 

\bigskip

\noindent{\bf Proof of Result 3:}   Consider user $i$ with value $V_i$ and suppose that all other users bid truthfully. Let $V_{K-1}$ and $V_K$ be the $K-1$st highest bid of others and the $K$th highest bid of others. If $V_i>V_{K-1}$ and $V_{K-1}>V_K$ then a bid by $i$ of $b_i$ such that $V_{K-1}>b_i>V_K$ gives $i$ a slot on the block at price $b_i$ while a truthful bid gives $i$ a slot on the block at price $V_{K-1}>b_i$. So truthful bidding is not a dominant strategy.

If a user's value is below $V_K$ there is no possible gain from bidding strategically as the price will be greater than the user's value for any bid. If $V_i\geq V_K$ then user $i$ would gain from a slot on the block. The price of that slot will be $V_{K-1}$ if $b_i \geq V_{K-1}$, as $V_{K-1}$ will be the lowest successful bid, and it will be $b_i$ if $V_{K-1}>b_i>V_K$, as in this case $b_i$ will be the lowest successful bid. So the gain that user $i$ can earn from strategic bidding (a non-truthful bid) is bounded by $(V_{K-1}-V_K)$, and this maximal gain can be earned only if $V_i\geq V_K$. Thus, user $i$'s expected gain from strategic bidding is bounded by $E[(V_{K-1}-V_K)]$ which converges to $0$ in the number of users.

\bigskip

\noindent{\bf Proof of Result 4:}   Follows immediately from an application of Swinkels [2001] and Jackson and Kremer [2006].

\bigskip

\newpage

\section{References}

\begin{description}
    \item Akbarpour, M. and and S. Li, 2018, Credible Mechanisms, \textit{Proceedings of the 2018 ACM Conference on Economics and Computation}, 371-371.
    \item Aune, R., A. Krellenstein, M. O’Hara, and O. Slama, 2017, Footprints on a Blockchain:  Trading and Information Leakage in Distributed Ledgers. \textit{Journal of Trading}, 12 (2), page 5-13.
    \item Bohme, R., N. Christin, B. Edelman,  and T. Moore, 2015. Bitcoin: Economics, Technology, and Governance, \textit{Journal of Economic Perspectives}, 29, 213-238.
    \item Cong, L., Z. He and J. Li, 2018, Decentralized Mining in Centralized Pools, SSRN,  https://ssrn.com/abstract=3143724.
	\item Easley, D. and J. Kleinberg, 2010, Networks, Crowds and Markets, \textit{Cambridge University Press}.
	\item Easley, D., M. O'Hara and S. Basu, 2017,  From Mining to Markets: The Evolution of Bitcoin Transactions Fees, \textit{Journal of Financial Economics}, forthcoming.
	\item Edelman, B. and M. Schwartz, 2010, Optimal Auction Design and Equilibrium Selection in Sponsored Search Auctions, \textit{American Economic Review}, May 2010, Vol. 100, No. 2: Pages 597-602.
	\item Eyal, I. and E. Sirer, 2014, Majority is Not Enough: Bitcoin Mining is Vulnerable. In \textit{Financial Cryptography and Data Security}, Christin, N., Safavi-Naini, R. (Eds.), Springer Heidelberg, Germany, pp. 436-454. 
	\item Gans, J. and H. Halaburda, 2015, Some Economics of Private Digital Currency. In Goldfarb, A., Greenstein, S.M., Tucker, C. E. (Eds.), \textit{Economic Analysis of the Digital Economy}, University of Chicago Press, Chicago, IL, Chapter 9. 
    \item Gandel, N. and H. Halaburda, 2016, Can We Predict the Winner in a Market with Network Effects?  Competition in cryptocurrency market, \textit{Games}, 7 (3), 1-21.
	\item Harvey, C., 2016, Cryptofinance, SSSRN, http://ssrn.com/abstract=2438299. 
	\item Huberman, G., J. Leshno and C. Moallemi, 2017, Monopoly Without a Monopolist: An Economic Analysis of the Bitcoin Pyament System, working paper, Columbia Business School.
	\item Houy, N., 2014, The Bitcoin Mining Game, available at SSRN, \\
	https://ssrn.com/abstract=2407834.
	\item Jackson, M. and I. Kremer, 2006, The Relevance of a Choice of Auction Format in a Competitive Environment, \textit{Review of Economic Studies}, Volume 73, Issue 4, Pages 961-981.
	\item Lavi, R., O. Sattath and A. Zohar, 2017, Redesigning Bitcoin's Fee Market, arXiv:1709.08881. 
	\item Ma, J., J. Gans and R. Tourky, 2018, Market Structure in Bitcoin Mining, NBER Working Paper No 24242.
	\item Malinova, K. and A. Park, 2016, Market Design with Blockchain Technology. SSRN, https://papers.ssrn.com/sol3/Delivery.cfm?abstractid=2785626.  
	\item Milgrom, R., The Economics of Competitive Bidding: A Selective Survey, in \textit{Social Goal and Social Organization: Essays in Memory of Elisha Pazner}, L. Hurwicz, D. Schmeidler and H. Sonnenschein (eds), Cambridge University Press, 261-289.
	\item Myerson, R., 1981, Optimal Auction Design, \textit{Mathematics of Operations Research},  Mathematics of Operations Research, Vol. 6, No. 1., pp. 58-73.
	\item Raskin, M. and D. Yermack, 2018, Digital Currencies, Decentralized Ledgers and the Future of Central Banking. In: Conti-Brown, P., Lastra, R., (Eds.), \textit{Research Handbook on Central Banking}, Edward Elgar, Publishing, 474-486.
    \item Raskin, M. and D. Yermack, 2017. Corporate Governance and Blockchains. \textit{Review of Finance}, 21, 7-31.
    \item Rosenfeld, M, 2011, Analysis of Bitcoin Pooled Mining Reward Systems, http://arxiv.org/pdf/1112.4980v1.pdf
	\item Swinkles, J., Efficiency of Large Private Auctions, 2001, \textit{Econometrica}, 69, 37-68. 
	\item Weber, R., 1983, Multi-Object Auctions, in \textit{Auctions, Bidding, and Contracting: Uses and Theory}, R. Engelbrecht-Wiggans, M. Shubik and R. Stark (eds), New York University Press, 165-191.
	\item Yao, A., 2018, An Incentive Analysis of some Bitcoin Fee Designs,	arXiv:1811.02351. 
	
\end{description}
\pagebreak

\pagebreak

\end{document}